\documentclass{article}
\usepackage{arxiv}

\usepackage[utf8]{inputenc} 
\usepackage[T1]{fontenc}    
\usepackage{hyperref}       
\usepackage{url}            
\usepackage{booktabs}       
\usepackage{amsfonts}       
\usepackage{amsmath}        
\usepackage{nicefrac}       
\usepackage{microtype}      
\usepackage{cleveref}       
\usepackage{graphicx}
\usepackage{natbib}
\usepackage{doi}
\usepackage{subcaption}
\usepackage{tabularx}
\usepackage{listings}
\usepackage{xcolor}

\hypersetup{hidelinks}

\providecommand{\Description}[1]{}

\definecolor{ndCodeBackground}{HTML}{272822}
\definecolor{ndCodeForeground}{HTML}{F8F8F2}
\definecolor{ndCodeKeyword}{HTML}{F92672}
\definecolor{ndCodeComment}{HTML}{75715E}
\definecolor{ndCodeString}{HTML}{E6DB74}
\definecolor{ndCodeAccent}{HTML}{66D9EF}
\definecolor{ndCodeFunction}{HTML}{A6E22E}
\definecolor{ndCodeBorder}{HTML}{49483E}

\lstdefinestyle{nd-base}{
  backgroundcolor=\color{ndCodeBackground},
  basicstyle=\ttfamily\scriptsize\color{ndCodeForeground},
  keywordstyle=\bfseries\color{ndCodeKeyword},
  commentstyle=\itshape\color{ndCodeComment},
  stringstyle=\color{ndCodeString},
  identifierstyle=\color{ndCodeForeground},
  frame=single,
  rulecolor=\color{ndCodeBorder},
  framerule=0.4pt,
  framesep=4pt,
  xleftmargin=1pt,
  xrightmargin=1pt,
  aboveskip=3pt,
  belowskip=3pt,
  breaklines=true,
  breakatwhitespace=false,
  columns=fullflexible,
  keepspaces=true,
  showstringspaces=false,
  tabsize=2
}
\lstdefinestyle{nd-python}{
  style=nd-base,
  language=Python,
  emph={np,torch,array,asarray,reshape,amax,exp,sum,range},
  emphstyle=\color{ndCodeFunction}
}
\lstdefinestyle{nd-fortran}{
  style=nd-base,
  language=Fortran,
  emph={transpose,reshape,matmul,sqrt,spread,maxval,exp,sum},
  emphstyle=\color{ndCodeFunction}
}
\lstdefinestyle{nd-output}{
  style=nd-base,
  language={},
  keywordstyle=\color{ndCodeAccent}
}

\title{NumericalDiff: Numerical Program Differencing via Value-Based Trace Alignment and Intervention}

\author{
  Xiang Chen \qquad Huanhuan Xia \\[0.4em]
  Zhongguancun Academy, Beijing, China \\
  Zhongguancun Institute of Artificial Intelligence, Beijing, China \\[0.4em]
  \texttt{chenxiang@bza.edu.cn} \qquad \texttt{huanhuanxia@zgci.ac.cn}
}

\renewcommand{\shorttitle}{NumericalDiff}

\begin{document}
\maketitle

\begin{abstract}
Numerical programs implementing the same computational task can exhibit different numerical behavior
because of their execution stacks, program representations, or program logic.  Source
text and final outputs alone provide limited diagnostic information: numerical
differences can propagate through corresponding computations, while distinct
executions can produce equal outputs.  We present NumericalDiff, an
execution-based method that records assignment-level numerical events, aligns
traces using a tolerance-aware longest common subsequence independent of
variable names and source locations, and optionally reexecutes after replacing
one intermediate value with its counterpart.  Among the original and all
single-intervention executions, it lexicographically maximizes matched events
and minimizes unmatched events.  NumericalDiff supports code comparison across language and
framework front ends, currently including NumPy, PyTorch, JAX, C, C++, Fortran, and Octave.
We evaluate NumericalDiff on 36 controlled cases comprising 675 pair/input instances: 36 execution-stack,
588 program-representation, and 51 program-logic instances.
All 24 operator-injection instances expose trace differences, including three
with tolerance-equal final outputs, while complete trace alignment is achieved for
all 441 canonical cross-representation instances.
Within the 51 program-logic instances, intervention
improves 24 alignments and raises macro-average matching coverage from 0.4210
to 0.6513.  The results show that value-based trace alignment and reexecution
can delimit propagated numerical differences across heterogeneous implementations.
\end{abstract}

\begin{figure*}[t]
  \centering
 \includegraphics[width=\textwidth]{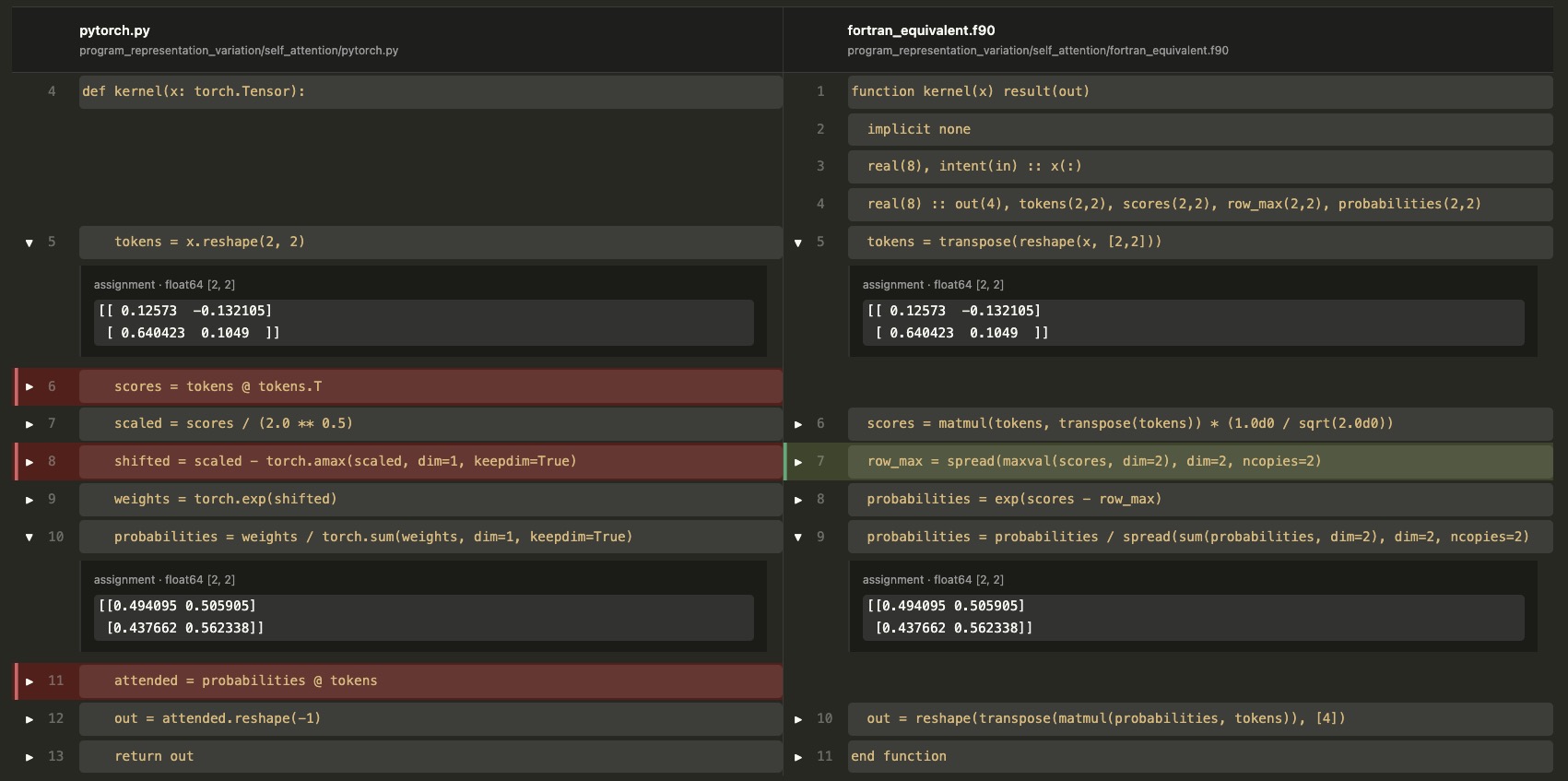}
  \caption{Demonstration of Code Comparison with NumericalDiff.}
  \Description{Demonstration of self-attention code comparison with NumericalDiff VS Code extension. On the left is a PyTorch implementation, and on the right is a Fortran implementation. The tool highlights the differences in the code and provides a side-by-side comparison of the two implementations.}
  \label{fig:teaser}
\end{figure*}

\section{Introduction}

Comparing two programs or two versions of a program is a recurrent task in
software engineering.  During testing and debugging, developers compare a
failing execution with a passing one to locate the first behavior that matters;
during maintenance and operation, they compare releases, configurations, and
deployments to diagnose regressions; and during migration, they compare an old
implementation with its replacement to determine whether behavior has been
preserved.  Existing techniques approach these tasks at several levels.
Source differencing identifies textual or syntactic changes, with modern tree
differencers providing fine-grained mappings between program elements
\citep{Falleri2024GumTree}.  Semantic differencing and differential symbolic
execution characterize behavioral changes beyond syntax
\citep{Jackson1994SemanticDiff,Person2008DSE}, while comparative execution,
slicing, and trace-based debugging explain how observed executions diverge
\citep{Sumner2013ComparativeCausality,Suzuki2017TraceDiff}.

Comparing numerical programs has become increasingly important as numerical
software is getting more widely deployed.  One prominent application
is accuracy alignment for deep-learning systems.  The same deep-learning
model may run on CPUs, GPUs, or accelerators; use different operator libraries,
runtimes, and numerical formats; or be implemented in
PyTorch, JAX, and other frameworks.  Such changes can preserve the intended
model while producing different intermediate tensors or outputs.  Cross-backend
validation has exposed and localized faults in deep-learning libraries
\citep{Pham2019CRADLE}, and differential testing continues to find discrepancies
across numerical libraries and heterogeneous platforms
\citep{Vanover2020FPDiff,Zhang2021HeteroFuzz}.  This need is especially visible
for large models, where a small local discrepancy may flow through many layers
before it becomes an observable loss of accuracy.

Another recurrent application is scientific-software migration, in which a
numerical program is reimplemented across languages such as Fortran, MATLAB,
C/C++, Python, or Julia while preserving its intended numerical semantics.
Migration can change the language, data layout, library calls, evaluation
order, and compiler behavior simultaneously.  Comparing only final outputs can
confirm that a test failed, but it does not reveal where two implementations
first became meaningfully different.  Relative debugging and numerical-drift
analysis therefore compare internal program states and, in some cases, replace
an observed value with a reference value to test whether later computation
reconverges \citep{Abramson2009RelativeIDE,Collins2013NumericalDrift}.

\begin{figure*}[tb]
  \centering
  \begin{subfigure}[t]{0.98\textwidth}
    \raggedright
    \begin{minipage}[t]{0.37\linewidth}
      \scriptsize\textbf{Shared source code}
\begin{lstlisting}[style=nd-python]
def newton(x):
    z = np.array(x, copy=True)
    for _ in range(60):
        f = z*z*z - 2*z + 2
        df = 3*z*z - 2
        z = z - f / df
    return z
\end{lstlisting}
    \end{minipage}\hfill
    \begin{minipage}[t]{0.29\linewidth}
      \scriptsize\textbf{float32 execution}
\begin{lstlisting}[style=nd-output]
x0 = -0.811595 -> 128.199707
... -> 1 -> 0 -> 1
cycling
\end{lstlisting}
    \end{minipage}\hfill
    \begin{minipage}[t]{0.29\linewidth}
      \scriptsize\textbf{float64 execution}
\begin{lstlisting}[style=nd-output]
x0 = -0.811595 -> 128.199149
... -> -1.769292
converged
\end{lstlisting}
    \end{minipage}
    \caption{The same Newton iteration code, cycling in \texttt{float32} but converging in \texttt{float64}.}
    \label{fig:demo_a}
  \end{subfigure}
  \par\medskip
  \begin{subfigure}[t]{0.98\textwidth}
    \raggedright
    \begin{minipage}[t]{0.48\linewidth}
      \scriptsize\textbf{PyTorch}
\begin{lstlisting}[style=nd-python]
def self_attention(x):
    tokens = x.reshape(2, 2)
    scores = tokens @ tokens.T
    scaled = scores / (2.0 ** 0.5)
    shifted = scaled - torch.amax(
        scaled, dim=1, keepdim=True)
    weights = torch.exp(shifted)
    probabilities = weights / torch.sum(
        weights, dim=1, keepdim=True)
    return probabilities @ tokens
\end{lstlisting}
    \end{minipage}\hfill
    \begin{minipage}[t]{0.48\linewidth}
      \scriptsize\textbf{Fortran}
\begin{lstlisting}[style=nd-fortran]
function self_attention(x) result(attended)
  real(8), intent(in) :: x(4)
  real(8) :: attended(2,2), tokens(2,2), &
      scores(2,2), scaled(2,2), shifted(2,2), &
      weights(2,2), probabilities(2,2)
  tokens = transpose(reshape(x, [2,2]))
  scores = matmul(tokens, transpose(tokens))
  scaled = scores / sqrt(2.0d0)
  shifted = scaled - spread(maxval( &
      scaled, dim=2), dim=2, ncopies=2)
  weights = exp(shifted)
  probabilities = weights / spread(sum( &
      weights, dim=2), dim=2, ncopies=2)
  attended = matmul(probabilities, tokens)
end function self_attention
\end{lstlisting}
    \end{minipage}
    \caption{The same self-attention algorithm, implemented in PyTorch and Fortran.}
    \label{fig:demo_b}
  \end{subfigure}
  \par\medskip
  \begin{subfigure}[t]{0.98\textwidth}
    \raggedright
    \begin{minipage}[t]{0.48\linewidth}
      \scriptsize\textbf{Iterative summation}
\begin{lstlisting}[style=nd-python]
def triangular_sum(n):
    total = 0
    for i in range(1, n + 1):
        total += i
    return total
\end{lstlisting}
    \end{minipage}\hfill
    \begin{minipage}[t]{0.48\linewidth}
      \scriptsize\textbf{Gauss's formula}
\begin{lstlisting}[style=nd-python]
def triangular_sum(n):
    return n * (n + 1) // 2
\end{lstlisting}
    \end{minipage}
    \caption{The same triangular sum task, implemented by iteration and Gauss's closed-form formula.}
    \label{fig:demo_c}
  \end{subfigure}
  \caption{Motivating examples.}
  \Description{Three text-only panels show the same Newton root-finding
    function cycling in float32 but converging in float64, equivalent
    self-attention functions written in PyTorch and Fortran, and equivalent
    triangular-sum functions using a Python loop and a closed-form formula.}
  \label{fig:demo}
\end{figure*}

Figure~\ref{fig:demo} demonstrates how numerical semantic differences are introduced
from different layers.  In \emph{execution-stack variation}, the same source code
runs with different hardware, drivers,
compilers, interpreters, runtimes, libraries,
numerical formats, or their configurations;
Figure~\ref{fig:demo_a} shows the same Newton iteration code cycling in
\texttt{float32} but converging in \texttt{float64}.  In
\emph{program-representation variation}, the same numerical algorithm is expressed in
different languages or frameworks; Figure~\ref{fig:demo_b} gives corresponding
PyTorch and Fortran self-attention implementations.  In \emph{program-logic
variation}, the same task is implemented with different algorithms; Figure~\ref{fig:demo_c}
contrasts iterative and closed-form triangular summation methods, which achieve the exactly same functionality but have different dynamic assignment structures.

Prior work addresses complementary parts of the comparison challenges
illustrated above.  Abstract semantic differencing reasons statically about
numerical relations between programs
\citep{partush2013abstract,delmas2019analysis}; cross-platform testers and
floating-point debuggers detect output inconsistencies or trace numerical error
\citep{Sawaya2017FLiT,SanchezStern2018Herbgrind,Guo2020pLiner}; and dynamic
comparison systems relate states through source correspondence, synchronized
control structure, inferred variable mappings, or domain-specific model layers
\citep{Gulwani2018Clara,Pham2019CRADLE,Suzuki2017TraceDiff}.  These constraints
are useful in their target settings but become restrictive when language,
framework, decomposition, or control flow differs: source locations and names
may not correspond, observed values require explicit shape and tolerance
policies, and loops or branches can change the event sequence.  Moreover, one
discrepant assignment can contaminate a long trace suffix, while merely locating
an unequal value does not establish whether replacing it would restore
downstream agreement.

We present \emph{NumericalDiff}, an execution-based differencing approach for
numerical functions.  We focus on pairs that serve the same purpose and have
similar computational logic; more distant algorithms are used only to
characterize boundary behavior.  For each input, NumericalDiff instruments
dynamically executed assignments, normalizes observed scalars, arrays, and
framework tensors, and computes a tolerant longest-common-subsequence alignment
without relying on variable or location identity.  To separate an originating
difference from propagation, it exhaustively evaluates the original executions
and every single assignment-value intervention in both directions: a value from
one execution is injected into the other, the recipient is reexecuted, and the
new trace is realigned.  NumericalDiff selects the candidate with the most
numerically matching events, breaking ties by fewer unmatched events.  When
later values reconverge, the diff presents the intervened assignment as the
localized difference and the suffix as conditional agreement.  Reexecution is
essential because an injected value can alter branches, loop trip counts, and
subsequent events.  This bounded objective tests whether one replacement
explains a propagated trace difference; it does not establish program
equivalence.  The implementation supports Python numerical frameworks, C, C++,
Fortran, and MATLAB-compatible programs executed by GNU Octave.

Table~\ref{tab:closest-work} compares NumericalDiff with the four closest
execution-based approaches.  \emph{Execution-based differencing} compares
behavior observed in concrete runs.  \emph{Runtime intervention} replaces a
value with one from the other execution and observes the resulting continuation.
\emph{Mapping-free event alignment} relates events through their observed
numerical values without requiring predefined correspondence between locations,
variables, or computational stages.  Collins et al. intervene at corresponding
instrumented locations; Clara infers a consistent variable mapping under
compatible control structure; CRADLE relies on a shared model-layer graph; and
TraceDiff compares update histories at corresponding relative positions.
NumericalDiff instead combines mapping-free event alignment with
intervention-conditioned trace differencing.

\begin{table}[tb]
  \caption{Comparison between the proposed method and closely related approaches.}
  \label{tab:closest-work}
  \centering
  \setlength{\tabcolsep}{4pt}
  \renewcommand{\arraystretch}{1.2}
  \begin{tabular}{@{}lccc@{}}
    \toprule
         & \shortstack{Execution-based\\differencing}
         & \shortstack{Runtime\\intervention}
         & \shortstack{Mapping-free\\event alignment} \\
    \midrule
    Collins et al.~\citep{Collins2013NumericalDrift}
      & $\checkmark$ & $\checkmark$ & $\times$ \\
    Clara~\citep{Gulwani2018Clara}
      & $\checkmark$ & $\times$ & $\times$ \\
    CRADLE~\citep{Pham2019CRADLE}
      & $\checkmark$ & $\times$ & $\times$ \\
    TraceDiff~\citep{Suzuki2017TraceDiff}
      & $\checkmark$ & $\times$ & $\times$ \\
    \midrule
    \textbf{NumericalDiff (ours)}
      & $\checkmark$ & $\checkmark$ & $\checkmark$ \\
    \bottomrule
  \end{tabular}
\end{table}

We summarize the contributions of this paper as follows:
\begin{itemize}
  \item We formulate numerical program comparison as execution-aware trace
    alignment with tolerant value agreement and value intervention,
    and define an exhaustive candidate-selection objective that separates a
    local difference from its propagated effects.
  \item We develop NumericalDiff, including a tolerant alignment
    and a single-intervention method that reexecutes programs to
    account for changed values and control flow.
  \item We implement a common observation and intervention protocol across
    Python, C, C++, Fortran, and MATLAB-compatible code.
  \item We build a benchmark spanning execution-stack,
    program-representation, and program-logic variation across AI,
    scientific computing, and general numerical programs.
\end{itemize}

\section{Methodology}

\subsection{Problem Formulation}
\label{sec:problem-formulation}

\paragraph{Scope and numerical observation points.}
Let $f_L,f_R:\mathcal{X}\to\mathcal{Y}$ be two numerical functions intended
to perform the same task through similar computational logic and a common
logical input/output interface.  Representation adapters, when needed,
translate this interface to each implementation.
For a supplied input $x\in\mathcal{X}$, we assume that each function executes
deterministically under its configured environment.  We observe numerical values
at supported assignment statements and at the terminal return.  Each dynamic
execution of an assignment constitutes a separate event; successive iterations
of the same statement produce distinct events.  An event
$e_i=(\ell_i,k_i,v_i)$ records its source location, kind
$k_i\in\{\mathsf{assign},\mathsf{return}\}$, and a snapshot of its numerical
value.  The value may be a scalar, an array, or an array-compatible framework
tensor.  An unmodified run produces
\begin{equation}
  T_f(x)=\langle e_1,\ldots,e_n\rangle.
\end{equation}
A terminal return is observed as an event but is not an intervention site.
Locations and variable names support reporting but do not constrain
cross-function event correspondence.

\paragraph{Numerical agreement and alignment.}
Let $C_\theta(u,v)\in\{0,1\}$ denote numerical agreement under a fixed
comparison policy $\theta$, which specifies representation normalization,
exact-shape compatibility, numerical tolerances, and handling of non-finite
values.  The current implementation does not match NaNs; the benchmark further
requires finite outputs and therefore does not exercise non-finite behavior.
We write $u\approx_\theta v$ when $C_\theta(u,v)=1$.  Two events may match only
when they have the same kind and their values agree.  An alignment $a$ consumes
both traces in execution order and consists of two-sided matches and one-sided
gaps.  Let $M(a)$ be its number of matches and $U(a)$ its number of gaps.
This observational agreement need not be transitive and does not establish
semantic equivalence.  Among the alignments for a fixed trace pair, tolerant
LCS maximizes $M(a)$.

\paragraph{Single-intervention candidate executions.}
An intervention replaces the value produced at one assignment event with a
value observed at an assignment event in the other execution.  The recipient
then continues from the injected value and is reexecuted from the original
input, because the replacement may change later values, branches, loop trip
counts, and trace length.  Arbitrary constants and return-value interventions
are excluded.  If an intervention were allowed directly at the output boundary,
copying the other execution's return value would trivially force equal outputs
without explaining any preceding difference.  Output agreement alone is
therefore not a useful objective, and return events are observation-only.  We
permit at most one assignment intervention because the present problem asks
whether one differing intermediate state can explain a propagated trace region;
this bound also makes exhaustive search possible.  Multiple independent origins
are outside the present formulation.

Let $A_L$ and $A_R$ be the assignment-event indexes in the original traces
$T_L=T_{f_L}(x)$ and $T_R=T_{f_R}(x)$.  For $p\in A_L$ and $q\in A_R$, let
$T_R^{p\to q}$ denote the right trace produced by replacing the value at right
event $q$ with the original value at left event $p$; define $T_L^{q\to p}$
symmetrically.  Only successfully completed, representable interventions are
admissible.  The candidate set is
\begin{equation}
  \mathcal{C}_x = \{(T_L,T_R)\}
    \cup \{(T_L,T_R^{p\to q})\mid p\in A_L,q\in A_R\}
    \cup \{(T_L^{q\to p},T_R)\mid p\in A_L,q\in A_R\}.
\end{equation}
For each $c\in\mathcal{C}_x$, let $a_c$ be a tolerant-LCS alignment and define
\begin{equation}
  \operatorname{score}(a_c)=\bigl(M(a_c),-U(a_c)\bigr).
  \label{eq:alignment-score}
\end{equation}
The optimization objective is
\begin{equation}
  c_x^* \in \operatorname*{arg\,max}_{c\in\mathcal{C}_x}
    \operatorname{score}(a_c),
  \label{eq:bounded-objective}
\end{equation}
where scores are compared in lexicographic order.  The objective therefore first
maximizes the number of tolerant matches and, among candidates with the same
number of matches, minimizes the number of unmatched events.  It is defined
independently for each supplied input over all zero- or one-intervention
candidate executions.  Agreement obtained after an intervention is conditional
on that intervention, and the objective does not establish equivalence on
unobserved inputs.

\subsection{Algorithm Design}
\label{sec:algorithm-design}

NumericalDiff compares the two functions independently for each supplied input.
It first executes both functions without intervention and obtains the event
traces $T_L$ and $T_R$ defined above.  Trace construction is execution driven:
only assignments that actually execute contribute assignment events, and all
events retain their runtime order.  The algorithm does not initially pair
variables, source locations, or syntactic statements across the two functions;
correspondence is determined from event kind, numerical agreement, and order.

\paragraph{Tolerant trace alignment.}
As a no-intervention baseline, we compute a longest common subsequence over the
two event sequences, replacing exact token equality with numerical agreement.
Let $L(i,j)$ be the maximum number of matches between prefixes of lengths $i$
and $j$, with $L(0,j)=L(i,0)=0$.  The recurrence is
\begin{equation}
L(i,j)=
\begin{cases}
L(i-1,j-1)+1,
  & e^L_i \text{ and } e^R_j \text{ agree under } C_\theta,\\
\max\{L(i-1,j),L(i,j-1)\}, & \text{otherwise}.
\end{cases}
\end{equation}
Backtracking yields an ordered sequence of matches and one-sided gaps.  A left
gap and a right gap are reported as unmatched events; they are not assumed to
be a semantic substitution merely because they are adjacent in the rendered
diff.  When multiple LCS alignments have the same length, the implementation
uses a fixed backtracking preference, making the selected alignment
deterministic for a fixed pair of traces.  This baseline can identify natural
reconvergence: a divergent block may be followed by matching events without
forcing the entire remaining suffix to be reported as different.

\paragraph{Single-intervention search.}
The intervention method exhaustively searches the bounded candidate space.
Let $A_L$ and $A_R$ be the dynamic indexes of assignment events in the two
traces; return observations are excluded because they do not affect
subsequent execution.  For every pair $(p,q)\in A_L\times A_R$, the algorithm
evaluates both directions.  In the $L\rightarrow R$ direction it reexecutes the
right function, replaces the value assigned at event $q$ with the
original value observed at left event $p$, and aligns the resulting right trace
against the unchanged left trace.  The $R\rightarrow L$ direction is symmetric.
The instrumentation records the recipient's originally computed value before
performing the replacement, while all later computation observes the replacement.
Consequently, the new alignment describes both the local discrepancy and the
continuation induced by the intervention.

Candidates are ranked by Equation~\eqref{eq:alignment-score}.
The baseline alignment is also a candidate.  A candidate replaces the current
best result only when its score is strictly larger, so ties retain the first
candidate in the fixed enumeration order.  The output records the selected
direction, recipient event, donor event, and resulting alignment.  This search
tests $2|A_L||A_R|$ intervened executions in addition to the original runs and
exactly optimizes Equation~\eqref{eq:bounded-objective}.  It does not compose
several interventions or optimize over that larger search space.

\paragraph{Loops.}
A loop is handled through its dynamic event occurrences rather than by a
special static loop-matching rule.  If one assignment in a loop body executes
$k$ times, it contributes $k$ distinct events with different dynamic indexes,
even though all carry the same source line.  LCS may therefore match individual
iterations, leave an extra iteration as a gap, or realign a common suffix after
different intermediate iterations.  The intervention search can target one
specific occurrence---for example, only the second execution of an assignment---
without modifying the other iterations of the same source statement.  The
recipient function is then rerun from its original input, so the replacement can
affect later loop-carried state and, when the loop bound or exit behavior depends
on that state, can also change the number of subsequent events.  NumericalDiff
does not require the two loops to have equal trip counts or identical loop
variables.  Loop-control bindings and predicates are not themselves observation
events unless their values are assigned at an otherwise supported observation
site.

\paragraph{Branches.}
For a conditional, the trace contains assignments from the path that was
actually taken; untaken branches contribute no placeholder events.  If the two
functions select different paths, their path-specific assignments appear as
unmatched regions.  Later events can nevertheless match when the paths
naturally reconverge to numerically agreeing states.  An intervention is not
restricted to preserving the original path: replacing an earlier assignment
may change a subsequent predicate, reach assignments absent from the original
trace, or skip assignments that were originally executed.  The algorithm
therefore reruns the recipient and aligns the newly produced trace rather than
patching a previously recorded sequence.  Branch predicates are not compared as
separate events unless explicitly materialized by an observed assignment; their
effect is represented by the resulting event sequence and values.

\paragraph{Result interpretation.}
NumericalDiff reports agreement between the original return values separately
from agreement in the selected trace alignment.
Any unmatched dynamic event makes the selected trace alignment non-equivalent,
but it does not by itself imply different return values.  Thus an extra
numerically neutral loop iteration or a branch that diverges and reconverges can
produce equal outputs while retaining a localized trace difference.  Conversely,
matches obtained after an intervention are reported together with that
intervention and should be interpreted as conditional agreement, not as
agreement of the original executions.

\paragraph{Complexity.}
Let $n$ and $m$ be the original trace lengths, $a=|A_L|$ and $b=|A_R|$ the
numbers of assignment events, and $R$ an upper bound on one instrumented execution.
The baseline tolerant LCS takes $O(nm)$ time and $O(nm)$ space.  The exhaustive
search evaluates $2ab$ interventions.  If every resulting trace has length at
most $N$, each candidate requires one reexecution and an $O(N^2)$ LCS, giving
$O\bigl((ab+1)R+abN^2+nm\bigr)$ time, including the original executions.
Since candidates are processed sequentially,
the peak auxiliary space is $O(N^2)$ in addition to the traces and backend
state.  When $a,b,n,m=O(N)$, the comparison component is $O(N^4)$ in the worst
case, plus $O(N^2R)$ reexecution cost.

\section{Implementation}

Before diving into the implementation details, we disclose that LLM-based coding agents, including Codex and Cursor, have assisted in implementing NumericalDiff, including language backends, comparison service, VS~Code extension, and in designing the benchmark cases. All LLM-generated content was reviewed and/or tested by the authors.

\subsection{Instrumentation Architecture}
\label{sec:trace-extraction}

NumericalDiff implements source-level dynamic instrumentation for Python, C,
C++, Fortran, and MATLAB-compatible \texttt{.m} files executed by GNU Octave.
The common entry point dispatches by file extension.  Python is transformed and
executed in process; the other languages use a prepared external runner that
parses the source, generates an instrumented program and runtime support, and
invokes the corresponding compiler or interpreter.  No external program is
translated into Python, and an unavailable tool or unsupported construct is
reported as an error rather than replaced by a fallback implementation.

Preprocessing is performed independently on the two sides: it identifies observation sites
inside the selected function but neither infers a trace from source order nor
constructs a cross-program mapping. Source locations are retained for the diff
view, while correspondence is established later from the values and dynamic
order of executed events. Across languages, each supported assignment follows
the same conceptual transformation: its right-hand side is evaluated once, the
result and dynamic event identity are recorded, and an optional donor value is
then assigned before execution continues.  Later statements therefore observe
the intervened state while the report retains the original value at the edited
site.  The event index denotes a dynamic occurrence rather than a source line;
loop iterations are therefore distinct, untaken branches produce no events,
and an intervention may change the subsequent control flow and trace length.
Variable names support source rewriting but are not alignment keys.

\subsection{Language-Specific Preprocessing}

Table~\ref{tab:language-instrumentation} summarizes the implemented front ends.
Each parser is used to locate syntactic write sites and source ranges; the
runtime value check remains authoritative because the tool does not attempt
whole-program numerical type inference.

\begin{table}[tb]
  \caption{Implemented source instrumentation backends.}
  \label{tab:language-instrumentation}
  \centering
  \begin{tabular}{@{}l@{\qquad}l@{}}
    \toprule
    Source & Parser / executor \\
    \midrule
    Python & \texttt{ast} / CPython \\
    C/C++ & Clang JSON AST / Clang \\
    Fortran & fparser2 / GFortran \\
    \texttt{.m} & tree-sitter / GNU Octave \\
    \bottomrule
  \end{tabular}
\end{table}

For Python, an \texttt{ast.NodeTransformer} rewrites the selected function and
preserves line numbers. Chained assignments produce one event per supported
direct-name target after the original statement. The transformed tree is
compiled in a fresh module namespace. Values are copied through NumPy;
PyTorch tensors are detached and moved to CPU first, while other array-compatible
objects, including JAX arrays, use NumPy conversion. Where supported, the
intervention adapter preserves PyTorch tensors, NumPy array dtypes, and scalar
types; JAX observations are supported through NumPy-compatible values.

For C and C++, Clang's JSON AST identifies one unambiguous top-level function
and byte ranges for source rewriting. The current interfaces are either
\texttt{double f(double)} or a flat one-dimensional array ABI with input,
output, and length arguments. The backend instruments initialized scalar
declarations and standalone \texttt{=}, \texttt{+=}, \texttt{-=},
\texttt{*=}, and \texttt{/=} statements. Array-element writes are observed as
separate scalar events on each dynamic execution. For canonical staged kernels,
initialized declarations of the bundled C \texttt{nd\_array} value type and
concrete Eigen dynamic arrays, vectors, and matrices are instead observed and
intervened on as whole values. This is type-directed instrumentation, not loop
recognition or event aggregation. The rewritten source is
combined with a small C runtime and driver and compiled by Clang or Clang++
using \texttt{-O0} and debug information.

For Fortran, fparser2 locates a single standalone function with one input and
an explicit \texttt{real(8)} or \texttt{double precision} result. Instrumented
whole-variable assignments call a generated generic \texttt{nd\_point}
procedure supporting ranks zero through two. A generated main program loads
the request and calls the function; GFortran compiles it with \texttt{-O0},
debug information, and runtime checks. For \texttt{.m} files, tree-sitter
locates one function with one named input and output and inserts
calls after whole-variable assignments. The transformed source and helper
functions are executed by GNU Octave.

Value-bearing returns are terminal observations rather than intervention
sites.  Python and scalar C/C++ evaluate the return expression once and record
it directly. Fortran and Octave record the result variable at explicit return
or normal function completion. The C/C++ array ABI records the output buffer
at fallthrough. Every successful external trace must end in such a return
event.

\subsection{Cross-Process Trace Protocol}

External backends exchange data with the Python controller through a small
text protocol. A request records the input rank, logical shape, values, and a
set of interventions represented by dynamic event index and donor value.
Inputs and donors are restricted to nonempty real scalar, vector, or matrix
values. Values are written with 17 significant decimal digits in logical
Fortran order; this makes transfer independent of the in-memory layout of the
originating NumPy array.

Each runtime emits one record containing source line, event kind, dtype, rank,
shape, and flattened values. The controller validates the record, reconstructs
its NumPy value in Fortran order, and normalizes all languages to a common
\texttt{TraceStep(line, value, kind)} representation containing a copied value.
Thus alignment and
intervention search consume one trace interface regardless of the source
language. Shape and dtype remain explicit, and numerical equality is evaluated
later under the request's absolute and relative tolerances. The runtime and
reader enforce a limit of 10,000 events and reject malformed traces, missing
terminal returns, and interventions that did not reach an executed assignment.

\subsection{Reexecution and Backend Lifecycle}

An external runner is prepared once per resolved source path and function name
within a comparison session. Generated source trees and executables live in a
session-scoped temporary directory. Compiled C, C++, and Fortran artifacts and
transformed Octave files are reused across the original executions and all
intervention candidates, while each rerun receives fresh request and trace
files and starts a fresh external process. The temporary directory is removed
when the comparison ends. Compilation and external execution have 60-second
and 20-second timeouts, respectively.

The same lifecycle supports both tolerant-LCS and brute-force intervention
alignment. The latter may repeatedly rerun either side with a donor value from
an assignment on the opposite trace.  Python replacements preserve the live
container and dtype where supported; external runtimes require compatible
rank and shape. Failures of individual speculative interventions are discarded
as invalid candidates, whereas missing tools, preprocessing failures, and
failed original executions are surfaced to the caller. Subprocesses and
timeouts provide lifecycle control, not a security sandbox; compared programs
must be trusted and may still perform external side effects.

\subsection{Current Coverage Limits}

The implementation observes selected source-level assignments, not every
memory write or mutation.  It does not recursively instrument helper functions,
recover writes hidden inside library calls, update all aliases after a
replacement, or support concurrent execution.  Python excludes
unpacking, attribute and subscript targets, assignment expressions, implicit
bindings such as loop targets, and bare returns. C/C++ excludes nested writes,
macro-expanded write sites, pointer targets, overloaded operators, and general
C++ objects. Fortran excludes modules, subroutines, pointers, derived types,
indexed targets, declaration initializers, and several compound statement
forms. Octave excludes indexed or multiple-output assignments, nested
functions, and global or persistent state. These restrictions define the
prototype's executable subset; they do not imply coverage of arbitrary
programs, optimized-binary behavior, or all side effects in the supported
languages.

\subsection{VS Code Extension Front End}

We expose NumericalDiff through a Visual Studio Code extension.  The command
\texttt{Numerical Diff: Compare Functions in Files} selects two visible editors
in left-to-right order, or exactly two open text-file tabs when that condition
does not hold.  It saves modified files, prompts for the function and, when
needed, input shape, and reads other comparison options from extension settings.
The extension invokes the same Python comparison service as other clients via a
JSON exchange over standard input and output.  Its Webview presents complete
function bodies side by side, places aligned events on shared rows, and marks
unmatched lines with the editor's removed and inserted colors.  Controls switch
samples, summarize and navigate differences, and report the selected algorithm.
Intervention-based results identify donor and
recipient lines and distinguish conditional matches from original agreement.

\section{Benchmark Construction}
\label{sec:benchmark-construction}

We instantiate the three variation layers of Figure~\ref{fig:demo} as 36
controlled cases from machine learning, scientific computing, and general
numerical programs.  Cases are labeled by the primary factor we change:
execution stack, program representation, or program logic.  A language or
framework migration can also change compilers, runtimes, and libraries, so
the label is not a claim that other layers are untouched.  Inputs, model
parameters, random seeds, and external state are fixed; each case has three
persisted inputs, an explicit numerical policy, and an independent output
reference.  The suite
yields 675 implementation-pair/input instances: 36 execution-stack, 588
program-representation (441 canonical comparisons and 147 cyclic alternative
probes), and 51 program-logic.  The cyclic probes compare alternatives across
ecosystems; mixed-factor edges also introduce a controlled program-logic
change rather than isolate representation variation.

The three categories answer complementary questions.  Execution-stack cases
ask whether a controlled environmental perturbation is visible in an
otherwise unchanged program.  Program-representation cases ask whether the
front ends preserve a common stage-level trace across languages and
frameworks.  Program-logic cases compare zero-intervention alignment with
bounded intervention search; exhaustive enumeration makes the selected
score optimal for Equation~\eqref{eq:bounded-objective}, but not a globally
minimum edit script, a unique causal explanation, or a proof of program
equivalence.  Across all three categories we report matching coverage
$2M/(2M+U)$ in $[0,1]$ to aggregate across trace lengths; it does not
replace the lexicographic objective.

\subsection{Execution-Stack Variation}

This category holds the abstract computation and program representation fixed
and varies only how that same program is executed.  Its potential sources are
hardware; numerical formats and arithmetic modes; operator or numerical
libraries; collective communication libraries; compilers and compilation
configurations; interpreters and their configurations; device drivers; and
runtimes and execution configurations.  A library change belongs here when it
occurs beneath an unchanged program interface; rewriting the program against a
different framework belongs to program-representation variation instead.

The current release instantiates four numerical-format cases and eight
controlled operator-library fault models.  Numerical-format cases use one
NumPy source file and execute each persisted input in both \texttt{float32}
and \texttt{float64}.  They cover rounding in stable Softmax, loss of small
updates during iterative accumulation, a precision-dependent Newton orbit,
and cancellation in a second-moment variance formula.  In particular, the
accumulation stagnates in \texttt{float32} but reaches its mathematical target
in \texttt{float64}; the Newton iteration enters a period-two orbit in
\texttt{float32} while converging to the real root in \texttt{float64}.

The operator cases likewise use one unchanged NumPy source and \texttt{float64}
inputs on both sides, but one execution profile deterministically biases the
result of \texttt{exp}, \texttt{matmul}, \texttt{max}, \texttt{sqrt},
\texttt{sin}, \texttt{divide}, \texttt{sum}, or \texttt{mean}.  These
profiles model the observable numerical consequence of a fault below an
unchanged program interface, independent of whether its physical origin is
hardware, a numerical library, a compiler, a driver, or a runtime.  They are
explicit harness-injected faults, not claims about real NumPy defects or
empirical tests of those physical components.  Each operator case has one
controlled changed execution component,
whereas a precision change can perturb many operations and need not have a
unique origin.  Both remain valid matching benchmarks: the former provides a
localized perturbation, while the latter measures how much agreement can be
recovered from distributed numerical drift.

\begin{table}[tb]
  \caption{Execution-stack variation cases. Each case executes one unchanged source file under two declared execution profiles.}
  \label{tab:benchmark-execution-stack}
  \centering
  \begin{tabularx}{\columnwidth}{@{}l l X@{}}
    \toprule
    Task & Domain & Numerical behavior exercised \\
    \midrule
    Stable softmax & AI & \texttt{float32} versus \texttt{float64} normalization \\
    Sigmoid operator fault & AI & \texttt{exp} result multiplied by $1.01$ \\
    Matrix operator fault & AI & \texttt{matmul} result multiplied by $1.005$ \\
    Maximum operator fault & AI & \texttt{max} result shifted by $0.01$ \\
    Iterative accumulation & Scientific & \texttt{float32} stagnation versus \texttt{float64} convergence \\
    Newton iteration & Scientific & \texttt{float32} period-two orbit versus \texttt{float64} convergence \\
    Square-root fault & Scientific & \texttt{sqrt} result multiplied by $1.002$ \\
    Sine operator fault & Scientific & \texttt{sin} result multiplied by $1.003$ \\
    Division operator fault & Scientific & \texttt{divide} result multiplied by $1.004$ \\
    Variance & General & \texttt{float32} versus \texttt{float64} cancellation in $E[x^2]-E[x]^2$ \\
    Reduction fault & General & \texttt{sum} result multiplied by $1.01$ \\
    Mean operator fault & General & \texttt{mean} result shifted by $0.02$ \\
    \bottomrule
  \end{tabularx}
\end{table}

\paragraph{Experimental protocol.}
Each case contains one unchanged \texttt{kernel.py} and three persisted inputs,
giving 36 execution-stack pair/input instances.  For a numerical-format case,
the two profiles differ only in array dtype.  Stable Softmax and variance use
shape-exact comparison with $\mathit{rtol}=10^{-10}$ and
$\mathit{atol}=10^{-12}$; Newton iteration uses $10^{-12}$ for both; and
iterative accumulation requires exact equality.  The accumulation and Newton
cases declare unequal outputs, whereas Softmax and variance measure whether a
given input exposes a difference rather than requiring one.

For an operator case, both profiles use \texttt{float64}; the faulty profile
intercepts exactly one named NumPy operator and applies the deterministic bias
shown in Table~\ref{tab:benchmark-execution-stack}.  All operator cases use
shape-exact comparison with $\mathit{rtol}=10^{-10}$ and
$\mathit{atol}=10^{-12}$.  Seven declare unequal outputs.  The maximum-reduction
case instead declares equal outputs: adding a constant to the maximum shifts
all Softmax logits equally and is removed by normalization, while the
intermediate execution remains different.  This case deliberately tests a
difference that final-output comparison cannot observe.

We run zero-intervention tolerant-LCS on every instance.  For workload
validation, the harness checks the unmodified profile against an independent
reference and checks the declared relation between the two outputs.  For the
differencing result, it records matched event pairs $M$, unmatched events $U$,
full-trace equivalence, and coverage $2M/(2M+U)$, then independently replays
every reported match against the full recorded values.  This experiment asks
whether a controlled execution-stack change is observable and how much of the
original traces remains aligned; it does not evaluate intervention recovery,
which is isolated in the program-logic experiment.

\subsection{Program-Representation Variation}

This category holds the abstract computation fixed while changing its program
representation.  Its primary sources are programming languages and numerical
or machine-learning frameworks.  The implementations must preserve the
mathematical definition, algorithmic stages, and intended data dependencies;
otherwise the case belongs to program-logic variation.  Each current task is
implemented in NumPy, PyTorch, JAX, C, C++, Fortran, and Octave, all using CPU
execution and nominal double precision.  For each task, a staged specification
defines the numerical value expected after every algorithmic stage, not only the
final result.  The seven canonical implementations realize this specification
with one whole-value assignment per stage.  Comparing all 21 unordered
canonical pairs on three persisted inputs yields 63 pure
program-representation instances per task and 441 across the seven tasks.

Each task additionally provides one alternative implementation in every
ecosystem.  These alternatives intentionally need not preserve the canonical
event sequence.  A semantics-preserving alternative computes the same output
through a different expression, control structure, or assignment granularity;
a faulty alternative also introduces a controlled program-logic change.  Each
task currently contains four semantics-preserving and three faulty alternatives.
To retain cross-representation comparisons without constructing an
uninformative alternative Cartesian product, we connect the alternatives in the
fixed cycle NumPy--PyTorch--JAX--C--C++--Fortran--Octave--NumPy.  Every edge
therefore crosses a language or framework boundary, and every alternative
selects one successor in another ecosystem.  No alternative is compared with a
canonical implementation in this part of the experiment.

The placement of the current labels around the cycle gives each task one edge
between two semantics-preserving alternatives and six edges between a
semantics-preserving and a faulty alternative; there is no faulty--faulty edge.
Across seven tasks, this produces seven equivalent--equivalent and 42
equivalent--faulty program pairs.  Running every pair on the three persisted
inputs yields 21 and 126 pair/input instances, respectively, or 147 alternative
instances in total.  Together with the 441 canonical instances, the category
contains 588 instances.  Faulty alternatives are identified by their output
classes, and edges containing one are treated as mixed-factor probes rather
than evidence isolating representation variation.

The cases preserve common stages rather than merely sharing a task name.  For
example, every canonical self-attention implementation explicitly computes the
score matrix, scaling, row-wise maximum subtraction, exponentiation,
normalization, and weighted aggregation.  NumPy, PyTorch, JAX, Fortran, and
Octave use their native array values; C uses the bundled \texttt{nd\_array}
value container; and C++ uses Eigen.  The latter two choices make C/C++ stages
observable and replaceable as complete values, yielding the same ordered event
granularity as the other five ecosystems.  This is a benchmark representation
contract rather than an analysis heuristic: the backend does not infer a
matrix or vector operation from a scalar loop.  Ordinary C/C++ element writes
therefore remain separate dynamic events, and loop-based alternatives are
explicit granularity challenges rather than members of the canonical
perfect-alignment oracle.  The self-attention alternatives also exercise
cross-language differences in expression granularity: the Fortran variant
materializes the broadcast row maximum as a separate assignment, whereas
other variants may fuse that value into the normalization expression.

Changing a representation can also change compilers, math libraries, and
runtime kernels.  The canonical comparisons consequently evaluate an
end-to-end migration under controlled device and precision settings, not the
isolated causal effect of syntax.  Their oracle is tolerant stage-by-stage
agreement with no unmatched events; bitwise identity is recorded separately
and is not required across ecosystems.

\begin{table}[tb]
  \caption{Program-representation variation cases. Every case has canonical and alternative implementations in NumPy, PyTorch, JAX, C, C++, Fortran, and Octave.}
  \label{tab:benchmark-program-representation}
  \centering
  \begin{tabularx}{\columnwidth}{@{}l l X@{}}
    \toprule
    Task & Domain & Fixed computation \\
    \midrule
    Sigmoid & AI & Element-wise $1/(1+\exp(-x))$ \\
    Squared-error loss & AI & Element-wise $0.5(x-0.25)^2$ \\
    Self-attention & AI & Two-token scaled dot-product attention with $Q=K=V$ \\
    Polynomial & Scientific & Horner evaluation of a fixed cubic polynomial \\
    Euler integration & Scientific & Two explicit steps for $y'=-0.5y$ \\
    Compound interest & General & Three compounding steps at a fixed one-percent rate \\
    Temperature conversion & General & Fahrenheit-to-Celsius affine conversion \\
    \bottomrule
  \end{tabularx}
\end{table}

\subsection{Program-Logic Variation}

This category changes the abstract computation while holding the execution
stack fixed---currently NumPy on CPU with \texttt{float64} inputs---and applies
each case's declared numerical policy consistently to both sides.  Its sources
are algorithmic choices and semantically
relevant code changes: mathematical reformulations, changed constants or
operators, altered data or control flow, inserted or removed events, and
different execution orders.  A source rewrite that preserves these properties
belongs to program-representation variation instead.  This category is designed
around trace behavior rather than application names.  It
contains examples in which a local discrepancy propagates, naturally
reconverges, is compensated by a later change, changes the executed branch or
loop, or alters event order without altering data dependence.  The 17 cases
contribute 51 pair/input instances.

Table~\ref{tab:benchmark-program-logic} summarizes the controlled program
differences.  Equal output does not imply equal traces: the
branch-reconvergence case follows different source paths before returning to
the same numerical state, and the operation-reordering case has crossing event
correspondences that a strictly sequential LCS cannot retain simultaneously.
The loop-occurrence case records repeated executions of one source assignment
and injects a fault only at the second dynamic occurrence.

Most cases closely match the intended setting of
Equation~\eqref{eq:bounded-objective}: the functions perform the same task with
similar logic, while a local change, branch, loop occurrence, or naturally
recovering state produces a controlled trace difference.  The remaining cases
broaden the difficulty range.  In particular, the two-rates case contains two
independent origins, the triangular-sum and alternative-algorithm cases have
more distant computational structures, and the operation-reordering case
exposes the order constraint of sequential LCS.  They still admit the same
quantitative evaluation: the achieved score and residual unmatched events show
how the current bounded search and sequential alignment behave when one
intervention cannot recover all agreement.

\begin{table}[tb]
  \caption{Program-logic variation cases and their differences.}
  \label{tab:benchmark-program-logic}
  \centering
  \setlength{\tabcolsep}{3pt}
  \begin{tabularx}{\columnwidth}{@{}l l X@{}}
    \toprule
    Task & Domain & Program difference \\
    \midrule
    Softmax & AI & Stable maximum-shifted versus direct exponentiation \\
    Attention scaling & AI & Divide scores by $\sqrt{d}$ versus by one \\
    ReLU recovery & AI & Different shifts collapse to the same zero region \\
    Sigmoid sign & AI & Exponentiate $-x$ versus $x$ \\
    Polynomial & Scientific & Horner evaluation versus expanded powers \\
    Euler sign & Scientific & Opposite sign in the first slope \\
    Stencil boundary & Scientific & Periodic versus zero-padded boundaries \\
    Threshold branch & Scientific & Different thresholds select different paths \\
    Loop occurrence & Scientific & Only the second execution of a repeated assignment is faulty \\
    Loop trip count & Scientific & One side executes an extra neutral iteration \\
    Variance & General & Centered two-pass versus second-moment formula \\
    Triangular sum & General & Iterative accumulation versus Gauss's closed form \\
    Two rates & General & Two independent rate constants are changed \\
    Compensating edits & General & A later change cancels an earlier change \\
    Extra assignment & General & One side computes an unused intermediate \\
    Branch reconvergence & General & Different branches reconverge after canonicalization \\
    Operation reordering & General & Two independent assignments execute in opposite orders \\
    \bottomrule
  \end{tabularx}
\end{table}

\paragraph{Experimental protocol.}
For every case, we compare its left and right NumPy functions on all three
persisted inputs, giving 51 pair/input instances.  We first run
zero-intervention tolerant-LCS and then run the exhaustive
single-intervention method with the same input and numerical policy.  Sixteen
cases use shape-exact comparison with $\mathit{rtol}=10^{-10}$ and
$\mathit{atol}=10^{-12}$; the integer-valued triangular-sum case requires exact
agreement.  All required outputs must be finite.  The intervention run includes the
zero-intervention alignment and enumerates both replacement directions for
every pair of dynamic assignment events; returns are observations but not
intervention sites.  Across the suite this gives 1,876 directional
intervention candidates, ranging from 12 to 408 per instance.

We validate the experiment at three levels.  First, independent references and
declared output relations check the original workloads.  Second, the harness
reexecutes the selected intervention, when any, and independently checks every
reported match against full runtime values.  It also checks annotated
single- or multiple-intervention witnesses, natural-recovery suffixes, loop
occurrences, and event-count or event-order properties where applicable.
Third, exhaustive enumeration establishes optimality only for the bounded
objective in Equation~\eqref{eq:bounded-objective}; it does not establish a
unique semantic alignment or causal explanation.

For each instance we record $(M_0,U_0)$ for tolerant-LCS and $(M^*,U^*)$ for
the selected zero- or one-intervention candidate, together with matching
coverage $C=2M/(2M+U)$, whether an intervention is selected, and comparison
time.  Coverage values are macro-averaged over pair/input instances; event
counts are summed.  We organize the cases into five disjoint behavior profiles:
single-change propagation, control-flow divergence, multiple origins,
different computational processes, and natural recovery or structural
variation.  We repeat each complete method run five times in fresh case-worker
processes.  Comparison time includes original traced executions, candidate
reexecutions, and alignment, but excludes process startup and independent
workload validation.  Discrete results must agree across repetitions; timing is
reported as the median.

Across all three categories, labels apply to the persisted inputs rather than
all possible inputs.  The suite is a controlled microbenchmark, not yet a
collection of independent real-world bugs.  It currently excludes broad input
distributions, nondeterminism, non-finite values, within-execution mixed
precision, and large applications.  Outside the staged canonical profile, it
also lacks automatic mappings between different event granularities and
independently adjudicated semantic correspondences for every case.  These
limitations are important when interpreting the later results: the benchmark
can establish workload reproducibility, replay soundness, bounded-objective
scores, matching improvement over a baseline, and behavior across controlled
difference patterns, but a high matching score does not by itself prove that
every correspondence is the uniquely intended one or establish external
validity for arbitrary numerical software.  Algorithm comparisons must
additionally use one method and policy uniformly; the per-group defaults used
for fast harness validation are not a fair cross-method evaluation.

\section{Evaluation and Analysis}
\label{sec:evaluation}

\subsection{Execution-Stack Variation}

All 12 cases satisfy their workload contracts on all three inputs, and all 36
zero-intervention alignments pass independent replay validation.  Thus every
match counted below is numerically valid under the case's shape and tolerance
policy.  Table~\ref{tab:execution-stack-results} separates the two controlled
variation families.

\begin{table}[tb]
  \caption{Execution-stack results under zero-intervention tolerant-LCS. Coverage is the macro-average of $2M/(2M+U)$ over pair/input instances.}
  \label{tab:execution-stack-results}
  \centering
  \begin{tabular}{lrrrrrr}
    \toprule
    Variation profile & $N$ & Output equal & Full trace & Matched $M$ & Unmatched $U$ & Coverage \\
    \midrule
    Numerical format & 12 & 2 & 2 & 36 & 1,146 & 0.2550 \\
    Operator injection & 24 & 3 & 0 & 33 & 168 & 0.2792 \\
    \midrule
    Overall & 36 & 5 & 2 & 69 & 1,314 & 0.2711 \\
    \bottomrule
  \end{tabular}
\end{table}

The eight operator models all have an observable dynamic effect: every one of
their 24 instances has a non-equivalent trace.  The \texttt{exp},
\texttt{matmul}, \texttt{sqrt}, \texttt{sin}, \texttt{divide}, \texttt{sum},
and \texttt{mean} injections also make all 21 corresponding outputs unequal.
In contrast, all three \texttt{max} instances retain tolerance-equal outputs
but have different maximum, shifted-logit, exponential, and reduction events.
Their final values are not bitwise identical.  This is the intended
process-different/result-equal control and shows concretely why output-only
comparison misses execution-stack differences.

Numerical-format behavior is input dependent rather than forced by the
harness.  Ten of its 12 instances have non-equivalent traces and unequal
outputs.  The remaining two are variance inputs whose \texttt{float32} and
\texttt{float64} executions are bitwise identical at every observed event;
they correctly remain fully matched.  The four cases therefore include both
exposed and benign format changes, while the stagnation and Newton cases
exercise qualitatively different long-running behavior rather than merely a
small last-bit perturbation.

Overall, only 5 of 36 final-output pairs are equal, but just 2 have fully
equivalent traces.  The gap is supplied by the three maximum-reduction
instances and demonstrates that output equality and trace equality answer
different questions.  The low macro-average coverage, 0.2711, is also
expected for value-based zero-intervention alignment: once a changed value
propagates, later assignments on the same source lines need not match.  The
especially long Newton traces contribute 1,052 unmatched events to the summed
count, so $U$ should not be read as 1,314 independent defects.  These results
validate that every injected operator fault takes effect and that the selected
format cases span both localized numerical drift and divergent iterative
dynamics; they are benchmark-characterization results, not evidence that a
single intervention can repair every execution-stack difference.

\subsection{Program-Representation Variation}

We use this category to test whether the implemented language front ends
can preserve a common numerical observation contract.  We run the tolerant-LCS
baseline with no intervention on all 441 canonical pair/input instances.  A
canonical instance is successful only if every assignment and return event is
matched in order under the case's shape-exact numerical policy, leaving no
event unmatched.  This is stronger than checking only the final output.  We
also replay every reported match against the recorded full values.

Table~\ref{tab:program-representation-results} reports the results.  All 441
canonical instances achieve complete trace equivalence: their 2,205 paired
events are matched and no event is left unmatched, giving matching coverage
1.0.  All final outputs are equal under the declared tolerance, while 425 of
441 are also bitwise identical.  The remaining 16 arise from ordinary
cross-ecosystem floating-point rounding; the largest observed absolute and
relative output differences are $8.88\times10^{-16}$ and
$2.22\times10^{-16}$, respectively.  Thus the tolerance admits only small
numerical differences here and does not conceal a missing stage or a structural
misalignment.  Each of the seven tasks independently obtains 63 of 63 complete
canonical alignments.

\begin{table}[tb]
  \caption{Program-representation results under zero-intervention tolerant-LCS. Coverage is the macro-average of $2M/(2M+U)$ over pair/input instances.}
  \label{tab:program-representation-results}
  \centering
  \begin{tabular}{lrrrr}
    \toprule
    Comparison profile & Instances & Output equal & Full trace & Coverage \\
    \midrule
    Canonical cross-representation & 441 & 441 & 441 & 1.0000 \\
    Equivalent--equivalent cyclic edge & 21 & 21 & 12 & 0.7429 \\
    Equivalent--faulty cyclic edge & 126 & 0 & 0 & 0.0577 \\
    \bottomrule
  \end{tabular}
\end{table}

The cyclic alternatives provide a separate cross-representation diagnostic;
none of these comparisons includes a canonical implementation.  The seven
equivalent--equivalent program pairs, each executed on three inputs, produce 21
pair/input instances.  All 21 have bitwise-identical outputs; 12 also have fully
equivalent traces, and their mean coverage is 0.7429.  The 42
equivalent--faulty program pairs similarly produce 126 instances.  All have the
declared unequal output, none has a fully equivalent trace, and their mean
coverage is 0.0577.  Alignment replay succeeds for all 588 canonical and
alternative instances, so every reported match is numerically valid under its
policy.  The first alternative group tests genuinely different but semantically
equivalent implementations across languages or frameworks; the second
intentionally combines representation and program-logic variation and is
interpreted only as a diagnostic probe.  In both counts, the factor of three is
the number of persisted inputs per program pair, not three additional
programs.

The result establishes the intended support claim for this controlled suite:
seven front ends can produce a common stage-level trace when the programs expose
the same stages.  It does not establish automatic semantic mapping for arbitrary
programs with incompatible event granularities.  In particular, a scalar C/C++
loop continues to produce one event per executed element assignment; no hidden
aggregation was used to obtain the canonical result.

\subsection{Program-Logic Variation}

All 17 cases pass their independent workload checks in all five repetitions,
and all 51 selected alignments pass replay validation.  Of the 51 original
pair/input instances, 30 have equal outputs and 21 have unequal outputs under
the declared policy.  These output counts characterize the workloads and are
not used as the alignment objective.

\begin{table}[tb]
  \caption{Program-logic results. $C_0$ and $C^*$ are macro-average matching coverage before and after the bounded intervention search. \emph{Improved} counts instances whose selected score is strictly greater than the zero-intervention score; \emph{Intervention} counts instances for which that candidate uses a value replacement.}
  \label{tab:program-logic-results}
  \centering
  \begin{tabular}{lrrrrrr}
    \toprule
    Behavior profile & $N$ & Output equal & Improved & Intervened & $C_0$ & $C^*$ \\
    \midrule
    Single-change propagation & 12 & 0 & 12 & 12 & 0.1190 & 0.8458 \\
    Control-flow divergence & 6 & 3 & 3 & 3 & 0.5429 & 0.6857 \\
    Multiple origins & 6 & 0 & 6 & 6 & 0.0000 & 0.2500 \\
    Different processes & 12 & 12 & 3 & 3 & 0.4539 & 0.5095 \\
    Recovery/structural variation & 15 & 15 & 0 & 0 & 0.7560 & 0.7560 \\
    \midrule
    Overall & 51 & 30 & 24 & 24 & 0.4210 & 0.6513 \\
    \bottomrule
  \end{tabular}
\end{table}

Table~\ref{tab:program-logic-results} shows that the search selects an
intervention exactly for the 24 instances in which it strictly improves the
lexicographic score; the other 27 retain the zero-intervention alignment.
Across all instances, the number of matched event pairs increases from 102 to
162 and unmatched events decrease from 385 to 265, giving a macro-average
coverage increase from 0.4210 to 0.6513.  Three attention-scaling instances
reach complete intervention-conditioned trace alignment; no original trace
pair is completely aligned.  The discrete scores and selected interventions
are identical in all five repetitions.

The behavior profiles clarify what the aggregate means.  All 12
single-change-propagation instances improve, from coverage 0.1190 to 0.8458,
because one replacement can restore a substantial downstream suffix.  The
loop-occurrence case confirms that the replacement can target only the second
dynamic execution of one source line.  In the control-flow profile, the three
threshold-branch instances improve after reexecution follows the intervened
state, whereas the three naturally reconvergent branch instances retain their
baseline.  The six multiple-origin instances improve only partially: one
replacement cannot repair both controlled origins.  Among the different-process
cases, the stable-versus-naive softmax instances improve, while polynomial,
variance, and triangular-sum implementations retain their baselines because
their event structures do not admit a better single replacement.  The recovery
or structural profile likewise selects no intervention; tolerant-LCS already
localizes the divergent region and retains naturally matching suffixes.

The median sum of measured comparison times over the 51 instances is
0.0700 seconds for tolerant-LCS and 2.1173 seconds for exhaustive intervention,
a $30.3\times$ increase.  The pooled per-instance medians are 1.21 and
9.04 milliseconds, respectively.  The difference between these ratios reflects
the skew from trace length: the triangular-sum loop generates the largest
candidate space and dominates aggregate intervention time while obtaining no
score improvement.  This result is consistent with the quartic worst-case
comparison cost and motivates a future guided search.  The experiment
demonstrates exact optimization of the present bounded objective and measurable
recovery of propagated agreement; it does not supply precision or recall
against a unique human correspondence oracle.

\section{Related Work}
\label{sec:related-work}

\subsection{Semantic Differencing and Relational Verification}

Program differencing ranges from edits over source structure to relations over
program behavior.  Structural techniques such as GumTree produce AST edit
scripts~\citep{Falleri2024GumTree}, whereas semantic differencing summarizes
changed dependencies or path behavior~\citep{Jackson1994SemanticDiff,Person2008DSE}.
Symbolic regression-verification systems encode two versions jointly and return
equivalence results or distinguishing executions; abstraction, refinement, and
partitioned classifications improve scalability and preserve partial results
when a complete proof is unavailable~\citep{Lahiri2012SymDiff,Badihi2020ARDiff,Glock2024PASDA}.
More generally, product programs combine two executions into one verification
problem, making the selected product construction an explicit account of how
the executions advance together~\citep{Barthe2011ProductPrograms}.
For numerical programs, relational abstract interpretation instead computes
sound abstractions of differences between corresponding variables.  DIZY and
the analyses of Delmas and Min{\'e} develop this line for numerical programs and
software patches~\citep{partush2013abstract,delmas2019analysis,Delmas2019Patches}.
All relational approaches must decide how executions correspond.  SCORE
searches bounded interleavings and favors steps that reduce the abstract
difference~\citep{Partush2014SCORE}; alignment-completeness results formalize
when such schedules suffice for a relational proof~\citep{Antonopoulos2023Alignment}.
For dynamic analyses, execution-point identifiers distinguish repeated
occurrences of a statement across executions~\citep{Sumner2013PEPID}.
NumericalDiff addresses the same correspondence problem at a different level:
it aligns concrete assignment events by their observed numerical values under a
tolerance, without requiring a joint syntax, common variable names, or a
predefined location map.

\subsection{Comparative Execution Diagnosis and Intervention}

Comparative debugging makes execution differences inspectable.  Relative
debugging, Scry, cozy, and TraceDiff relate divergent states or value histories
to program context~\citep{Abramson2009RelativeIDE,Burg2015Scry,Helbling2025Cozy,Suzuki2017TraceDiff},
while Dual Execution and LDX couple executions to handle environmental or
control-flow divergence~\citep{Kim2015DualExecution,Kwon2016LDX}.  Dependency
analyses compare control flow, correct and failing implementations, or paired
slices~\citep{Qi2009DARWIN,Banerjee2010Golden,Weeratunge2010DualSlicing}; InPreSS
and CoReX further reduce propagated differences while retaining explanatory
context~\citep{Badihi2023InPreSS,Badihi2026CoReX}.
Intervention-based debugging modifies state and observes the outcome.  State
delta debugging, cause transitions, value replacement, and Comparative
Causality progressively isolate relevant values and moments
~\citep{Zeller2002CauseEffect,Cleve2005Causes,Jeffrey2008ValueReplacement,Jeffrey2009MultipleFaults,Sumner2013ComparativeCausality}.
For numerical drift, Collins et al. overwrite differing values at matched
instrumentation points~\citep{Collins2013NumericalDrift}; pLiner, Ciel, and
multi-level FLiT instead vary precision or compilation regions to localize
environment-induced variability~\citep{Guo2020pLiner,Miao2023Ciel,Bentley2019MultiLevel}.
These approaches generally assume shared locations or the same source, whereas
NumericalDiff searches value correspondences between independently implemented
functions.
Reference executions can also guide source repair: Clara and SARFGEN minimize
repairs after dynamic or structural alignment~\citep{Gulwani2018Clara,Wang2018SARFGEN},
whereas AutoRNP derives numerical approximations from mathematical
structure~\citep{Yi2019AutoRNP}.  NumericalDiff repairs neither source nor a
global edit sequence.  It ranks zero- and single-assignment-intervention trace
alignments, so restored agreement is input-specific, intervention-conditioned
evidence rather than proof of a unique cause.

\subsection{Numerical Debugging and Differential Testing}

Numerical debuggers compare an execution with a high-precision or perturbed
shadow.  FPDebug and Herbgrind track error propagation, Verificarlo samples
rounding behavior, and NUMFL ranks error-associated expressions
~\citep{Benz2012FPDebug,SanchezStern2018Herbgrind,Denis2016Verificarlo,Bai2017NUMFL}.
They diagnose one implementation against a precision or statistical oracle;
NumericalDiff instead compares peer implementations without designating either
as ground truth.
Differential testing compares libraries and environments.  FPDiff tests
synonymous numerical functions, FLiT and HeteroFuzz expose platform or compiler
variability, and CRADLE, DeepREL, and Predoo target deep-learning backends, API
relations, and operator precision
~\citep{Vanover2020FPDiff,Sawaya2017FLiT,Zhang2021HeteroFuzz,Pham2019CRADLE,Deng2022DeepREL,Zhang2021Predoo}.
These systems find inconsistent executions but do not combine assignment-trace
alignment with counterpart-value intervention.
Input generation is complementary: shadow symbolic execution and hybrid
fuzzing distinguish program versions~\citep{Palikareva2016Shadow,Noller2020HyDiff},
while search-based testing, FPGen, and FPCC seek large floating-point errors
~\citep{Zou2015Inaccuracies,Guo2020FPGen,Yi2024FPCC}.  Because higher precision
is not always a valid oracle~\citep{Wang2016PrecisionSpecific}, NumericalDiff
makes inputs and tolerances explicit policies and leaves defect classification
outside the alignment result.

\section{Conclusion}

This paper presented NumericalDiff, an execution-based approach for comparing
same-purpose numerical functions across execution stacks, program
representations, and program logic.  It aligns assignment-level values using
tolerance-aware LCS and searches zero- and single-intervention executions for
the alignment with the greatest numerical agreement.  Across 36 controlled
cases and 675 pair/input instances, NumericalDiff completely aligns canonical
cross-representation programs, exposes operator faults hidden by equal final
outputs, and recovers substantial downstream agreement after propagated
program-logic differences.  These results show that intermediate-value
comparison complements source- and output-level differencing.  Future work includes guided intervention search, input value design, and more efficient cross-representation alignment.

\subsubsection*{Acknowledgments}
This work was supported by Zhongguancun Academy under Grant No. XTS0030.

\bibliographystyle{unsrtnat}
\bibliography{references}

\end{document}